\documentclass[%
 reprint,
superscriptaddress,
 amsmath,amssymb,
 aps, prl,
]{revtex4-2}

\usepackage{graphicx}
\usepackage{dcolumn}
\usepackage{bm}
\usepackage{soul}
\usepackage{hyperref}
\usepackage{comment}

\usepackage{color}

\usepackage{textcomp}

\newcommand{\DIYfancyhead}{\hspace{18mm} FERMILAB-PUB-24-0725-SQMS-T-TD}

\begin{document}
\DIYfancyhead

\title{An Improved Bound on Nonlinear Quantum Mechanics using a 
Cryogenic Radio Frequency Experiment}

\author{Oleksandr Melnychuk}
\email[Correspondence to: ]{melnit@fnal.gov}
\author{Bianca Giaccone}
\email[Correspondence to: ]{giaccone@fnal.gov}
\affiliation{Fermi National Accelerator Laboratory, Batavia, IL 60510, USA}
\author{Nicholas Bornman}
\affiliation{Fermi National Accelerator Laboratory, Batavia, IL 60510, USA}
\author{Raphael Cervantes}
\affiliation{Fermi National Accelerator Laboratory, Batavia, IL 60510, USA}
\author{Anna Grassellino}
\affiliation{Fermi National Accelerator Laboratory, Batavia, IL 60510, USA}
\author{Roni Harnik}
\affiliation{Fermi National Accelerator Laboratory, Batavia, IL 60510, USA}
\author{David E.~Kaplan}
\affiliation{Department of Physics \& Astronomy, The Johns Hopkins University, Baltimore, MD  21218, USA}
\author{Geev Nahal}
\affiliation{Fermi National Accelerator Laboratory, Batavia, IL 60510, USA}
\author{Roman Pilipenko}
\affiliation{Fermi National Accelerator Laboratory, Batavia, IL 60510, USA}
\author{Sam Posen}
\affiliation{Fermi National Accelerator Laboratory, Batavia, IL 60510, USA}
\author{Surjeet Rajendran}
\affiliation{Department of Physics \& Astronomy, The Johns Hopkins University, Baltimore, MD  21218, USA}

\author{Alexander O. Sushkov}
\affiliation{Department of Physics, Boston University, Boston, Massachusetts 02215, USA}
\affiliation{Department of Electrical and Computer Engineering, Boston University, Boston, MA 02215, USA}
\affiliation{Photonics Center, Boston University, Boston, MA 02215, USA}
\affiliation{Department of Physics \& Astronomy, The Johns Hopkins University, Baltimore, MD  21218, USA}

\date{\today}
             
\begin{abstract}
There are strong arguments that quantum mechanics may be nonlinear in its dynamics. A discovery of nonlinearity would hint at a novel understanding of the interplay between gravity and quantum field theory, for example. As such, experiments searching for potential nonlinear effects in the electromagnetic sector are important. Here we outline such an experiment, consisting of a stream of random bits (which were generated using \texttt{Rigetti}'s \texttt{Aspen-M-3} chip) as input to an RF signal generator coupled to a cryogenic detector. Projective measurements of the qubit state, which is originally prepared in an equal superposition, serve as the random binary output of a signal generator. 
Thereafter, spectral analysis of the RF detector would yield a detectable excess signal predicted to arise from such a nonlinear effect. A comparison between the projective measurements of the quantum bits vs the classical baseline showed no power excess. This sets a new limit on the electromagnetic nonlinearity parameter $|\epsilon| \lessapprox 1.15 \times 10^{-12}$, at a 90.0\% confidence level. This is the most stringent limit on nonlinear quantum mechanics thus far and an improvement by nearly a factor of 50 over the previous experimental limit. 
\end{abstract}

\maketitle

\section{\label{sec:intro}Introduction}

Linearity of time evolution is a fundamental axiom of quantum mechanics. However, in physical systems linearity is typically an approximation to a more complete description of the system. The full evolution of the system can be viewed in terms of a power series, where the linear terms are observationally prominent with nonlinear terms suppressed by energy scales. Recent theoretical work~\cite{kaplan2022causal}, building on prior insights~\cite{Kibble:1978vm, Polchinski:1990py}, has shown that a similar structure is possible in the time evolution of quantum mechanics itself. It has been shown that there is a broad class of consistent (causal and gauge invariant) extensions of quantum mechanics where the time evolution of the system is nonlinear. In this class of theories, one can view the time evolution equation of a quantum field theory as a power series expansion in the quantum states. The term that is linear in the quantum state is simply the first term in this expansion and is observationally dominant. This results in linear evolution emerging as the leading order approximation of a more complex time evolution. 

The main observation of Ref.~\cite{kaplan2022causal} is that the Hamiltonian that describes the time evolution of a quantum state $|\Psi\rangle$ of a quantum field theory can be made state-dependent -- thus leading to nonlinear time evolution -- by shifting various bosonic field operators in the Hamiltonian by a small amount proportional to the expectation value of the operator in the full quantum state $|\Psi\rangle$. In quantum electrodynamics, these nonlinear terms effectively shift the conventional Lagrangian $A_{\mu}J^{\mu}$ to the term $\left(A_{\mu} + \epsilon \langle \Psi| A_{\mu} | \Psi\rangle \right)J^{\mu}$. Here $A_{\mu}$ is the electromagnetic 4-vector potential, $J^{\mu}$ a 4-vector current and $\epsilon$ parametrizes the magnitude of the nonlinear term. Thus, the current $J^{\mu}$ effectively sees a modified 4-vector potential $\left(A_{\mu} + \epsilon \langle \Psi| A_{\mu} | \Psi\rangle \right)$.  It was shown in Ref.~\cite{kaplan2022causal} that despite over a century of precise quantum mechanical measurements (for example, in atomic, optical, nuclear and collider experiments), observational limits on causal nonlinear quantum mechanical evolution were weak, but that nevertheless, potent experimental probes of such nonlinearities are possible with dedicated experiments. 

\begin{figure*}[t]
    \centering
    \includegraphics[width=0.8\linewidth]{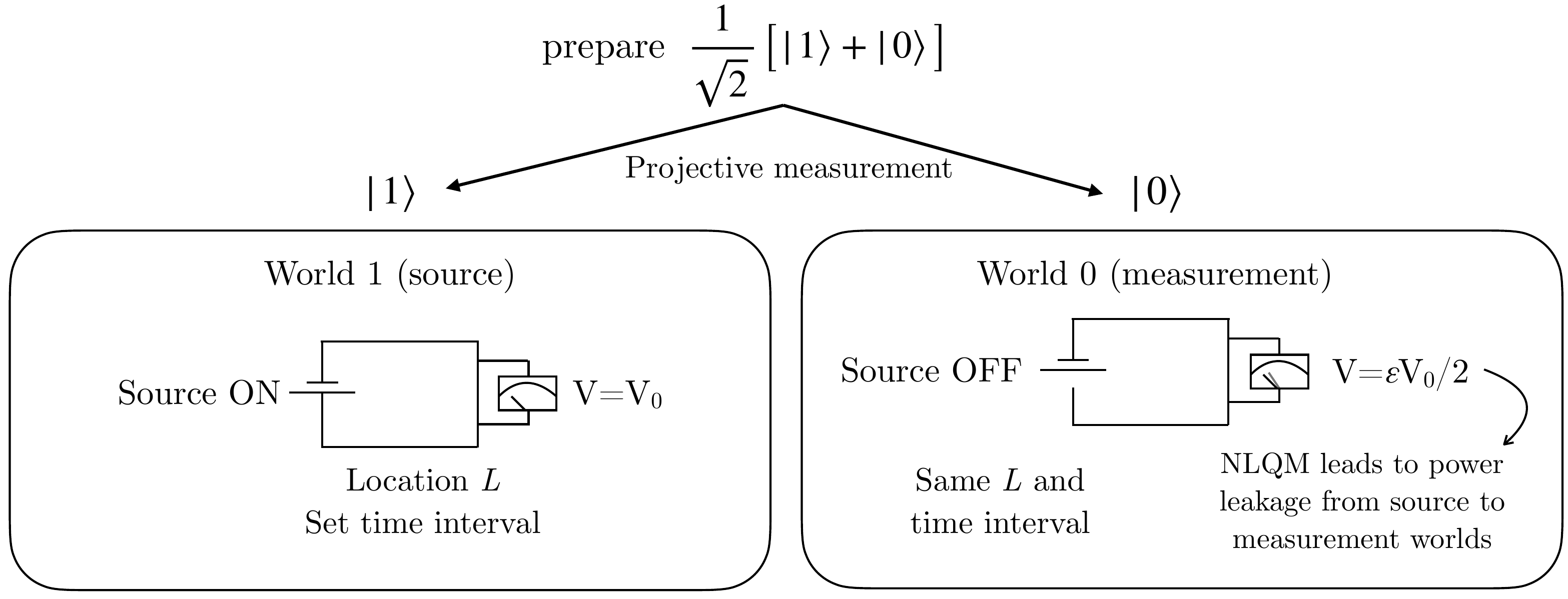}
    \caption{A schematic of our test of nonlinear quantum mechanics (NLQM). A qubit is prepared in a superposition state and measured, creating a macroscopic superposition, two worlds. In the world in which the qubit is measured to be in the $|1\rangle$ state an electric circuit is powered with a voltage of $V_0$ during a predetermined time. In the world in which the qubit is measured to be $|0\rangle$, the source is kept off, but the voltage of the same circuit is measured to high precision in the same predetermined time. Because the expectation value of the voltage, averaged over both worlds is $V_0/2$, NLQM will effectively lead to a leakage of voltage of size $\epsilon V_0/2$ from the source world to the measurement world.}
    \label{fig:cartoon}
\end{figure*}

The experimental protocols to test Ref.~\cite{kaplan2022causal} relied on the following fact: nonlinear evolution leads to interactions of the quantum state with itself. This manifests itself in the form of interactions between different ``branches" or ``arms'' of the wave function, also referred to as ``worlds". Each branch corresponds to one possible outcome of the measurement of the quantum system, and all branches coexist after the measurement~\cite{Everett_evModPhys.29.454}. Based on this, there are two kinds of experiments that can probe nonlinear quantum mechanics. First, one can consider interferometric experiments where a quantum system is placed in a superposition and anomalous interactions between the arms of the superposition manifest as a phase shift in the interferometer. This protocol was implemented in an ion interferometer~\cite{Broz:2022aea} which searched for the Coulomb field of one arm of the ion interferometer interacting with the other arm. In these interferometric experiments, it is essential to maintain quantum coherence of the superposition. 

Surprisingly, causal nonlinear quantum mechanics leads can lead to effects that persist even when the arms of the superposition get entangled with an environment and decohere. This phenomenon, dubbed the ``Everett Phone''~\cite{Polchinski:1990py}, gives rise to a second class of potent experimental probes. 
A search relying on this is sketched schematically in Figure~\ref{fig:cartoon}.
First, a macroscopic quantum superposition is created. Since the superposition need not be coherent, it is easily created by performing a measurement on a quantum system such as a spin or a qubit. The many-worlds interpretation suggests that this measurement, which is simply an interaction between the quantum system and the measuring device, creates a macroscopic superposition of two worlds: one in which the qubit was measured to be $|0\rangle$ and another where it was measured to be $|1\rangle$. At a fixed location $L$ in the laboratory, we turn on a voltage $V_0$ if the qubit is measured in $|1\rangle$. In the case of $|0\rangle$ , we use a voltmeter to measure the voltage signal with no voltage source turned on at location $L$. 
The Schr{\"o}dinger equation predicts that the full quantum state $|\Psi\rangle$ is the superposition of both these worlds, and thus it can be verified that $\langle \Psi | A_{\mu} | \Psi\rangle$ is non-zero at $L$. In the presence of the nonlinear term $\epsilon \langle \Psi| A_{\mu} | \Psi\rangle J^{\mu}$, the current $J^{\mu}$ will respond to an additional voltage $\epsilon V_0/2$ irrespective of the qubit outcome that the current is entangled with. That is, the voltage that was turned on in the world when the qubit was $|1\rangle$ leaks into the world where it is $|0\rangle$. This phenomenon, a robust consequence of any causal nonlinear extension of quantum mechanics~\cite{Polchinski:1990py, kaplan2022causal}, is thus once again simply the interaction between two arms of a superposition that nonlinearities would enable. However, its persistence, even in the presence of decoherence, could enable much more potent experimental probes of this scenario than is possible with quantum coherent interferometric probes, as in Ref.~\cite{Broz:2022aea}. The first experiment implementing this protocol was performed in Ref.~\cite{polkovnikov2023experimental}, obtaining a bound of $|\epsilon| \lessapprox 4.7 \times 10^{-11}$. 

The purpose of the present paper is to report on an experiment that implemented this protocol using a high-power radio frequency (RF) setup, which improved the bound to $|\epsilon| \lessapprox 1.15 \times 10^{-12}$. Here again, we measure a qubit system. If a $|1\rangle$ measurement is obtained, we turned on an RF source with applied power $P_A$ at a fixed location $L$ in the laboratory. If $|0\rangle$ was obtained, we turned on a device that measured RF power without any RF source being turned on. In the presence of the nonlinear term $\epsilon \langle \Psi| A_{\mu} | \Psi\rangle$, the detectors will experience an additional power $\epsilon^2 P_{A}/4$ irrespective of the qubit outcome they are entangled with. Thus, in the world where the qubit was in the 0 state and there was no power source, the detector will see RF power $\epsilon^2 P_{A}/4$. If $P_M$ is the smallest power that can be experimentally measured, then the corresponding value of $\epsilon$ that the experiment is sensitive to is given by
\begin{align}
|\epsilon_{\textrm{lim}}| = 2 \times \sqrt{P_{M}/P_{A}}
\label{eqn:epsilon}
\end{align}

\section{\label{sec:procedure}Experimental Procedure and Setup}

The experiment described in this publication follows the same general logic that was developed in Ref.~\cite{polkovnikov2023experimental}. Distinct states of the two-level quantum system (superconducting transmon qubits in our case~\cite{rothTransmon2023}) are made to be uniquely correlated with the two specific sets of actions in the macroscopic experimental setup. Then, the macroscopic effects, which may be possible only in nonlinear quantum mechanics, are sought.

\begin{figure}[h!]
\includegraphics[scale=0.5]{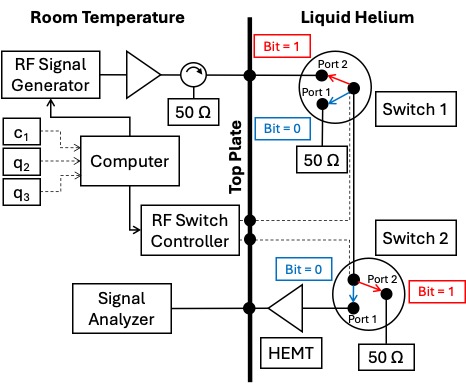}
    \caption{Schematic of the experiment including both room temperature (RT) and 1.9\,K portion. The top plate indicates the border between the RT and the cold environment. The connections between the RT and cold parts happen at the top plate. $c_1$, $q_2$, and $q_3$ are used to indicate the classical and the two quantum bit strings, which are used as input for the experiment. Based on the bit value (0 or 1), the configuration of the switches is set as indicated in the figure and the RF signal generator is turned on (if {bit~=~1}) or maintained off (if {bit~=~0}). The load connected to port 1 of switch 1 is used in the {bit~=~0} configuration to terminate the cryogenic amplifier input and prevent the introduction of RT noise. The high-power load connected to port 2 of switch 2 is used in the {bit~=~1} configuration to safely discard the high RF power coming from the RF source, to avoid saturating and potentially damaging the cryogenic amplifier.}
    \label{fig:ExperimentSchematic}
\end{figure}
The schematic of the setup is shown in Fig.~\ref{fig:ExperimentSchematic}. The main components are the RF generator, room temperature (RT) high-power (HP) amplifier, cryogenic amplifier (HEMT), Signal Analyzer (SA), the system of two RF switches and the main computer that automates the experiment. Then the power spectrum is recorded on the Signal Analyzer.

\begin{figure}[h!]
   \includegraphics[width=0.45\textwidth]{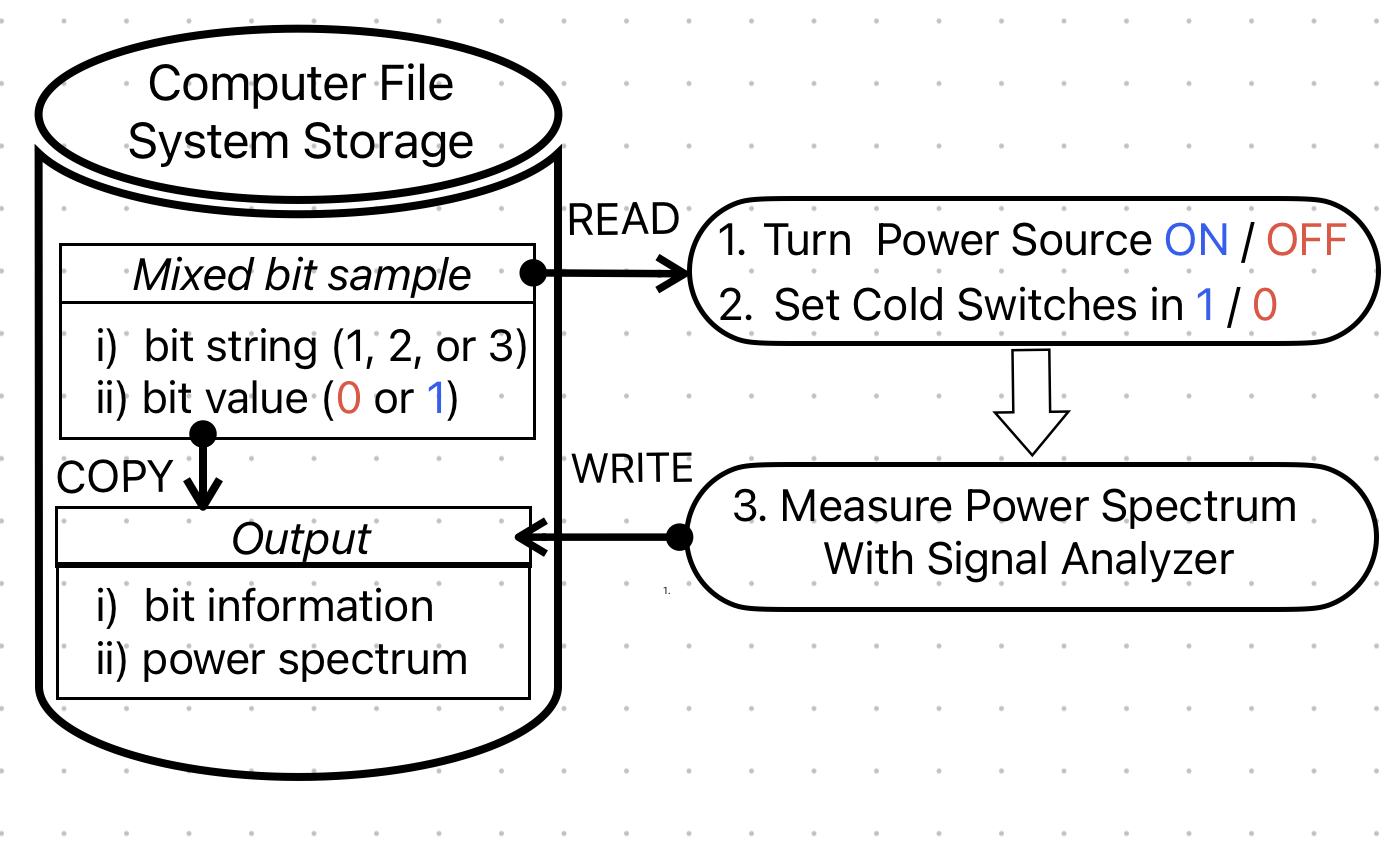}
    \caption{Logical layout of the \texttt{python} script running the experiment from the main computer.}
    \label{fig:PythonScriptSchematic}
\end{figure}

The circuit has two modes of operation. Depending on the value of each read bit (0 or 1), the main computer turned on or off the power source and set the cryogenic RF switches according to the schematic. In the {bit~=~0} case, the source is off and the SA records the power spectrum of the amplified thermal noise plus the amplified nonlinearity signal. In the {bit~=~1} case, the source is on and the amplified power from the source is directed to a matched HP load (connected to port 2 of switch 2). The control is executed with the \texttt{python} script starting with reading the bits, one at a time from the mixed sample, which contains both quantum and classical bits. One bit corresponded to one experimental data point. The logical layout of the \texttt{python} script is shown in Fig.~\ref{fig:PythonScriptSchematic}.

The two bit values (0 and 1) define the two circuit configurations of Fig.~\ref{fig:ExperimentSchematic} that correspond to the two branches of the total wave function, which contains both possibilities at the same time.  In the {bit~=~0}  branch the source is turned off, the switches are set so that there is no electrical connection between the source signal path and the readout path, and the input of the cryogenic amplifier is terminated by a matched load, connected to port 1 of switch 1. In case of {bit~=~1}, the source is turned on, the switches are connected so that the signal from the source reaches the central port of switch 2. The nonlinear quantum mechanics (NLQM) theory predicts that signal from the {bit~=~1} branch can leak to the {bit~=~0} branch. Relative to this schematic, a certain fraction of the signal would leak from the {bit~=~1} branch to the {bit~=~0} branch at the central port of switch 1, and at the cable connecting the two switches, thus producing a signal excess readable through the cryogenic amplifier.

The experiment was designed to be run by a computer according to a pre-defined algorithm to exclude the possibility of creating undesired ``quantum dilution'' phenomenon described in Ref.~\cite{kaplan2022causal} so that the actions of the experimenters should not have any correlations with the states of the qubit. In addition, the experimenters were not allowed to know the values of the individual bits at any point in time. In advance of the experiment, three bit sequences (samples) were created using one classical and two quantum random number generators. The quantum samples were obtained using two qubits on the 8-qubit \texttt{Aspen-M-3} quantum processor available on \texttt{Rigetti}'s Quantum Cloud Services~\cite{RigettiREF}. Then, the bits from the three samples were randomly mixed into one ``mixed sample'' and the same ``dilution safety'' rules were followed as for the individual bit samples.

In order to be sensitive to the nonlinear effects, the emission of the signal in one branch and the measurement of the leaked signal in the other branch needed to overlap in time. To achieve appropriate experimental sensitivity and to avoid ``quantum dilution'', it was crucial to ensure that the sequence of actions in the experiment in the two branches was properly synchronized. Whenever the measurement of the signal in one branch was performed, the source in the other branch had to be turned on for the duration of the measurement. This was achieved by proper timing of each action in the branch where the experiment was performed. Dedicated studies were performed to time the duration of each measurement step and ensure that the total duration of the {bit~=~1} case coincided with the duration of the {bit~=~0} operations in order to ensure synchronization between the {bit~=~1} and {bit~=~0} branches.

The information on potential nonlinear effects is contained in the data recorded using the quantum sample bits with the value of 0. This sample was used in the physics analysis. Classical data obtained from {bit~=~0} were used as control samples to assess noise level, as in Ref.~\cite{polkovnikov2023experimental}. For the rest of this paper, we will refer to the bits obtained from the quantum processor as quantum bits, and data obtained with the quantum samples as quantum data. The same nomenclature applies to the classical case.

\section{\label{sec:equipment}Equipment}
The RF signal was generated by a Rohde and Schwarz 8\,kHz -- 20\,GHz SMA100B signal generator and amplified at room temperature by an OPHIR 5265 0.7-4.2\,GHz, 200\,W RF high-power amplifier. RF-Lambda 2.1 – 2.9\,GHz RFC26-200 wideband high-power circulator was used to protect the high-power amplifier from high reflected signal. On the readout side, the signal was amplified by the cryogenic low noise HEMT amplifier LNF-LNC2\_4A from Low Noise Factory (S/N 061H). AC-DC switching adaptor MW GP25A13D and LNF-PS\_3 power supply designed for LNF low noise cryogenic amplifiers were used to power the HEMT. Rohde and Schwarz FSW-26 2\,Hz -- 26.5\,GHz Signal and Spectrum Analyzer was used for recording the power spectrum.
Radial R583F33121RF RF Switches operated via National Instruments CDAQ-9178 module with two NI 9477 cards RF Switch Controller were used for switching the circuit between {bit~=~0}, {bit~=~1} and calibration configuration. Keithley 2231A-30-3 triple channel DC power supply was used for powering the controller. All equipment was controlled in a synchronized manner from the main (MacBook Pro laptop) computer except for the switch controller for which an auxiliary (Windows 10 DELL laptop) computer was used. All devices were connected through a local network via a Netgear ProSafe GS108 Gigabit Ethernet switch. SMA100B signal generator, Keysight ENA E5080B, 9\,kHz-9\,GHz Network Analyzer and Keysight EPM Series N1914A Power Meter with Keysight E-Series E9300A average power sensor were used for cable calibrations.
The cryogenic section of the experiment circuit was placed inside a liquid helium dewar at the cavity vertical test facility in Fermilab's Applied Physics and Superconducting Technology Directorate (APS-TD). During the NLQM experiment data taking, the temperature in the dewar remained between 1.915\,K and 1.927\,K. 

\section{\label{sec:quantumbit}Generation of Bits}

Quantum samples were generated on \texttt{Rigetti}'s~\cite{RigettiREF} hardware according to the following procedure:
\begin{enumerate}
\item Initialize the qubits in their ground states via active reset
\item Apply a Hadamard gate (H-gate) to each qubit, which sets each one's state to be an equal superposition of their ground and first excited states in the $Z$ basis
\item Read out the individual qubit states simultaneously, and write the bit value to a classical register
\item Repeat steps 1-3 for a sufficient number of shots

\end{enumerate}
There are three types of fidelities associated with this process, namely the average single qubit gate fidelity obtained via randomized benchmarking one-gate operation fidelity (relevant for the H-gate) ($F_{1QRB}$), the classical readout fidelity ($F_C$), and the active reset fidelity.
Quantum bits, with true randomness arising from a quantum process, were generated with two qubits, named qubits 101 and 102, one qubit at a time. The $F_{1QRB}$ fidelity and active reset fidelity of both qubits were above 99\%. We used $F_{1QRB}$ fidelity as an estimate of the H-gate fidelity $F_{H}$. Classical readout fidelities were 98.3\% and 86.1\% for qubits 101 and 102, respectively. Fidelities as well as other qubit parameters, are calibrated and measured by \texttt{Rigetti} routinely every six hours. The quoted numbers come from the measurement that took place within the 6 hours of the generation of the bits. The classical bits were produced with a pseudo-random generator implemented in \texttt{random} package from the standard \texttt{python} library~\cite{pythonrandom}. The two quantum-generated bits samples from qubits 101 and 102 were then mixed with the classical bits sample to produce a trackable (but blinded) random mix to be used in the experiment.

\section{\label{sec:canalysis}Calibration and Limit Extraction Procedure}
The use of classical, or pseudo-random, bits serves a dual scope in the experiment. The pseudo-random bits act as control sample to assess the systematics of the experiment and extract the noise level, similarly to Ref. \cite{polkovnikov2023experimental}. In addition, the classical bits with value 0 are also used to validate the analysis, having the possibility to fully unblind this subset of data without incurring in ``quantum dilution''.

\begin{figure}[h!]
\includegraphics[width=0.49\textwidth]{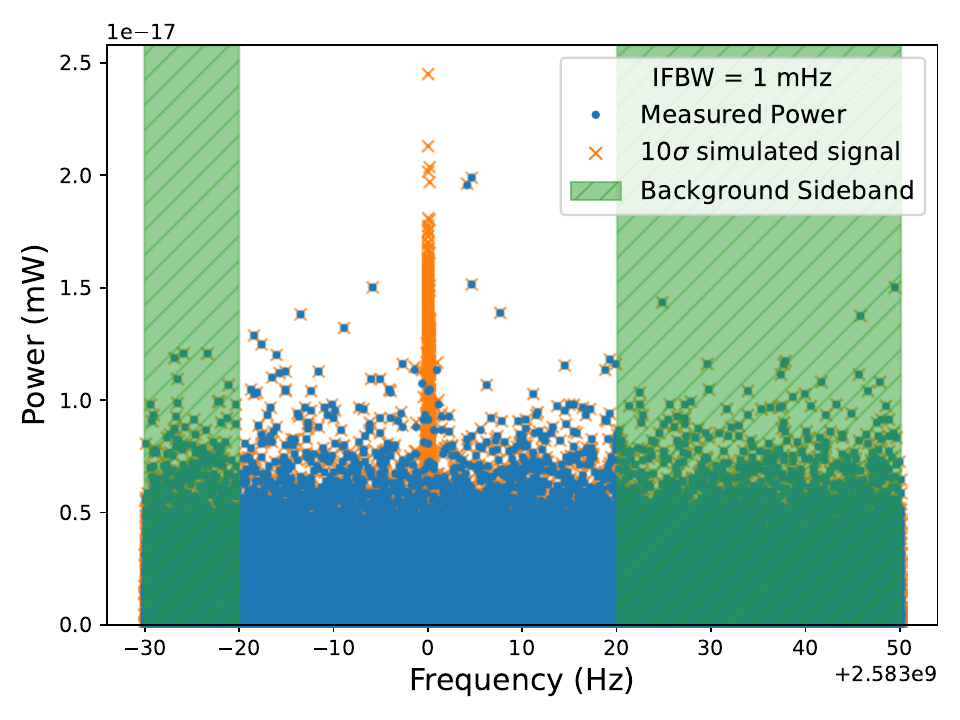}
    \caption{Power measured by the signal analyzer for the {bit~=~0} configuration in Fig.~\ref{fig:ExperimentSchematic}. The green shading represents the background sidebands used to estimate the background. The orange points represent a simulated 10-sigma nonlinearity signal.} 
    \label{fig:sascreenshot}
\end{figure}

\begin{figure}[h!]
    \centering
\includegraphics[width=0.5\textwidth]{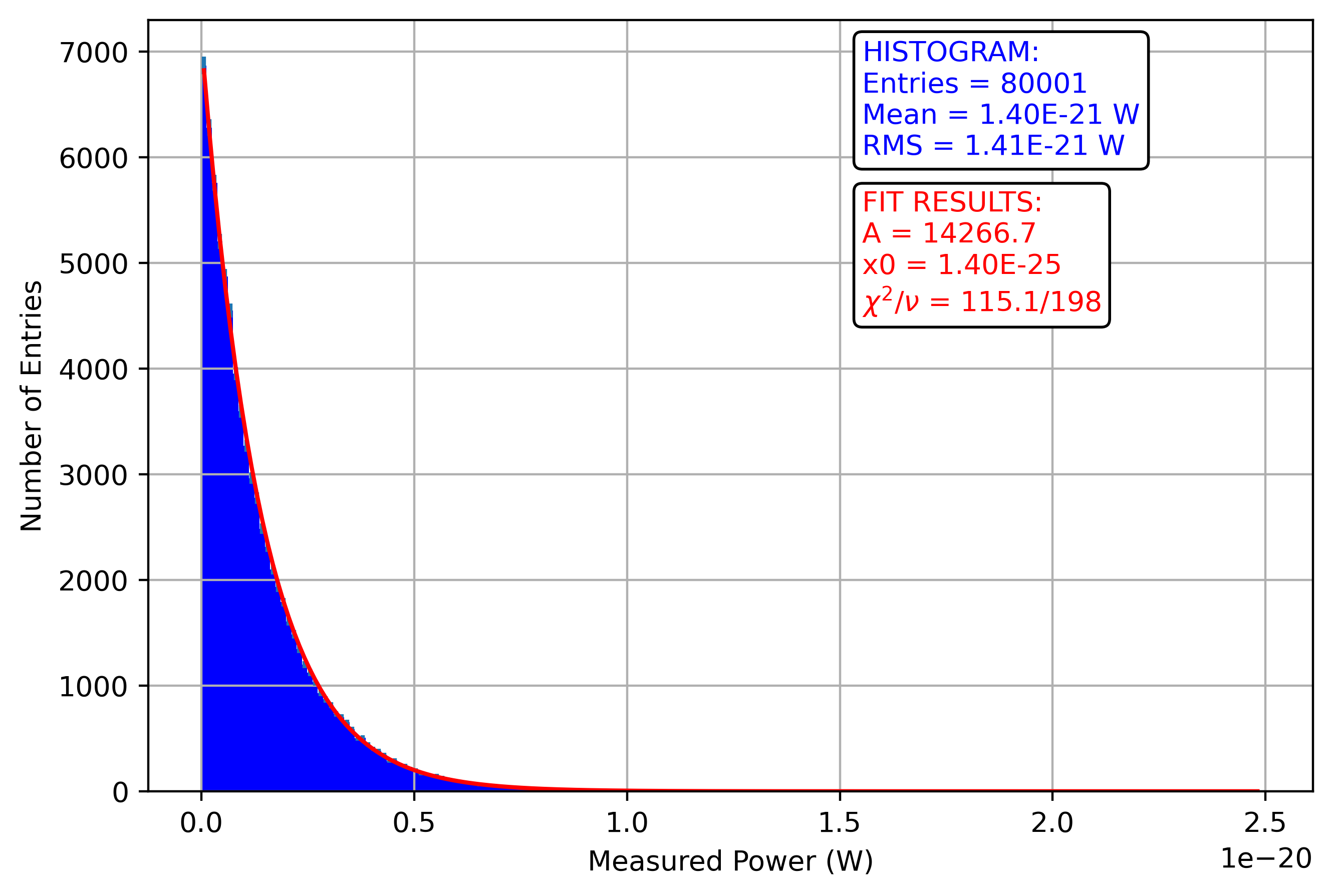}  
    \caption{Measured power histogram and the fit for the data shown in Fig.~\ref{fig:sascreenshot} }
    \label{fig:sascreenshothistoandfit}
\end{figure}

During the experiment 66 mixed bits were processed, of which 25 were classical, of which 10 bits had the value equal to 0. These 10 classical data points were analyzed first and a fictitious quantum mechanics (QM) electromagnetic (EM) nonlinearity limit was derived. This fictitious limit was derived, as an exercise, under the assumption that these data were recorded with random samples that were quantum in nature (while they were in fact sampled classically), obtained from qubits with imperfect fidelities, and simply serves as a reference to check the analysis protocol. This is necessary since the data obtained with the quantum samples were kept blinded to minimize risks of ``quantum dilution''. Fig.~\ref{fig:sascreenshot} shows a single power spectrum measurement for classical {bit~=~0} case. Power in the signal region, i.e., power measured in the 1\,mHz interval around the source frequency will be referred to as $P_s$. The areas highlighted as background sidebands are used to estimate the background mean and $\sigma$. For each bit, one individual power spectrum is acquired. As a result, due to the lack of averaging, the power distribution follows a chi-square probability density with 2 degrees of freedom. This is shown in Fig.~\ref{fig:sascreenshothistoandfit}, which represents the histogram of the power spectrum acquired for one bit, fitted using a rescaled two degrees of freedom $\chi^2$ distribution fit function of the form $\chi^{2}(x) = A \times \frac{1}{2} \times e^{ - (x - x_0)/2}$, where  A and $x_0$ are free parameters of the fit.

The analysis was performed on the calibrated power spectrum at the HEMT input.

\subsection{Calibration of Cables}
Calibration of cables, the circulator, RF connections, and switches was carried out to measure the insertion loss introduced by these components. Cable calibrations were performed using Vector Network Analyzer (VNA) and Power Meter (PM) measurements of different sections of the circuit in Fig.~\ref{fig:ExperimentSchematic}. The measurements were performed at room temperature. Cable calibration error in this type of experimental setup is typically 5\%~\cite{melnychuk2014error}. To be conservative in the limit-setting, the error was considered to be i) 10\% instead, ii) 100\% systematic in nature and iii) one-sided, i.e., increasing the $P_M$  and decreasing the $P_A$. The difference between RF cable losses at room temperature and cryogenic temperature is within the error. Cable attenuation between the high-power amplifier output and the high-power load was 7.10\,dB. Cable attenuation between HEMT output and SA input was 1.89\,dB.
 
\subsection{Calibration of High-Power Amplifier}
From measurements of input power and output power, 200\,W amplifier gain was found to be 60.73 $\pm$ 0.6\,dB. To be conservative in the limit-setting, the error was considered to be 100\% systematic in nature, one-sided and decreasing the nonlinearity (NL) signal, i.e., 60.73\,dB - 0.6\,dB = 60.13\,dB gain factor was used for calibrating the $P_A$.

\subsection{Calibration of Cryogenic Low Noise HEMT Amplifier} 
HEMT calibration was performed by combining the measurements of amplified thermal noise and amplified generator signal. During the thermal noise measurement, both switches were put in the {bit~=~0} state (Fig.~\ref{fig:ExperimentSchematic}), the source was turned off, the high-power amplifier was turned off and the HEMT amplifier was turned on. The amplified 1.921\,K thermal noise was measured to be \mbox{-154.6\,dBm $\pm$ 0.1\,dB} (fit error). SA was used with resolution bandwidth (RBW) of 1\,Hz and pre-amplifier gain of 30\,dB. Using this measurement and Eqn.~(\ref{eqn:TNvsGain}) a relationship between two unknown quantities, noise temperature $T^{Noise}_{HEMT}$ and gain $G_{HEMT}$, was established: 
\begin{align}
 S(f)=\frac{k_b \times b \times (T_{dewar} + T^{Noise}_{HEMT}) \times G_{HEMT}}{ IL } 
  \label{eqn:TNvsGain},
\end{align}
where $S(f)$ indicates the measured power spectrum in units of W/Hz, $k_b$ is the Boltzmann constant, $b$ is the frequency bin size, and $IL$ is the insertion loss after the HEMT in linear units.

The HEMT gain was obtained from a separate measurement of the amplified -130\,dBm RF generator signal. During the measurement switch 1 and switch 2 were put in {bit~=~1} and {bit~=~0} state respectively. The generator output was connected via a 5-foot RF cable to port 1 of the circulator. The OPHIR 200W amplifier was bypassed in this measurement. The generated signal was much larger than both the expected HEMT effective noise and the ambient thermal noise. Both of these could be ignored and a simple equation $P^{HEMT}_{out} = P^{HEMT}_{in} \times G_{HEMT}$ was used to obtain the HEMT gain. Here $P^{HEMT}_{in}$ is the power output from the signal generator calibrated for the insertion loss of the RF chain between the signal generator and the HEMT input (equal to 7.53\,dB). The combination of the two measurements enabled us to obtain the HEMT gain and noise temperature:
\begin{align}
T^{Noise}_{HEMT} = \text{\small 4.108\,K}\\
G_{HEMT} = \text{\small 38.087\,dB}
\end{align}
Note that the attenuation factor of the circuit section from the HEMT output to the SA input cancels out between the HEMT calibration and the data calibration. This was verified both analytically and by performing both the HEMT and the data calibrations using several fictitious values of the attenuation factor and obtaining the same calibrated data.

\subsection{Power Limit}
\begin{figure}[h!]
    \centering
    \includegraphics[width=0.5\textwidth]{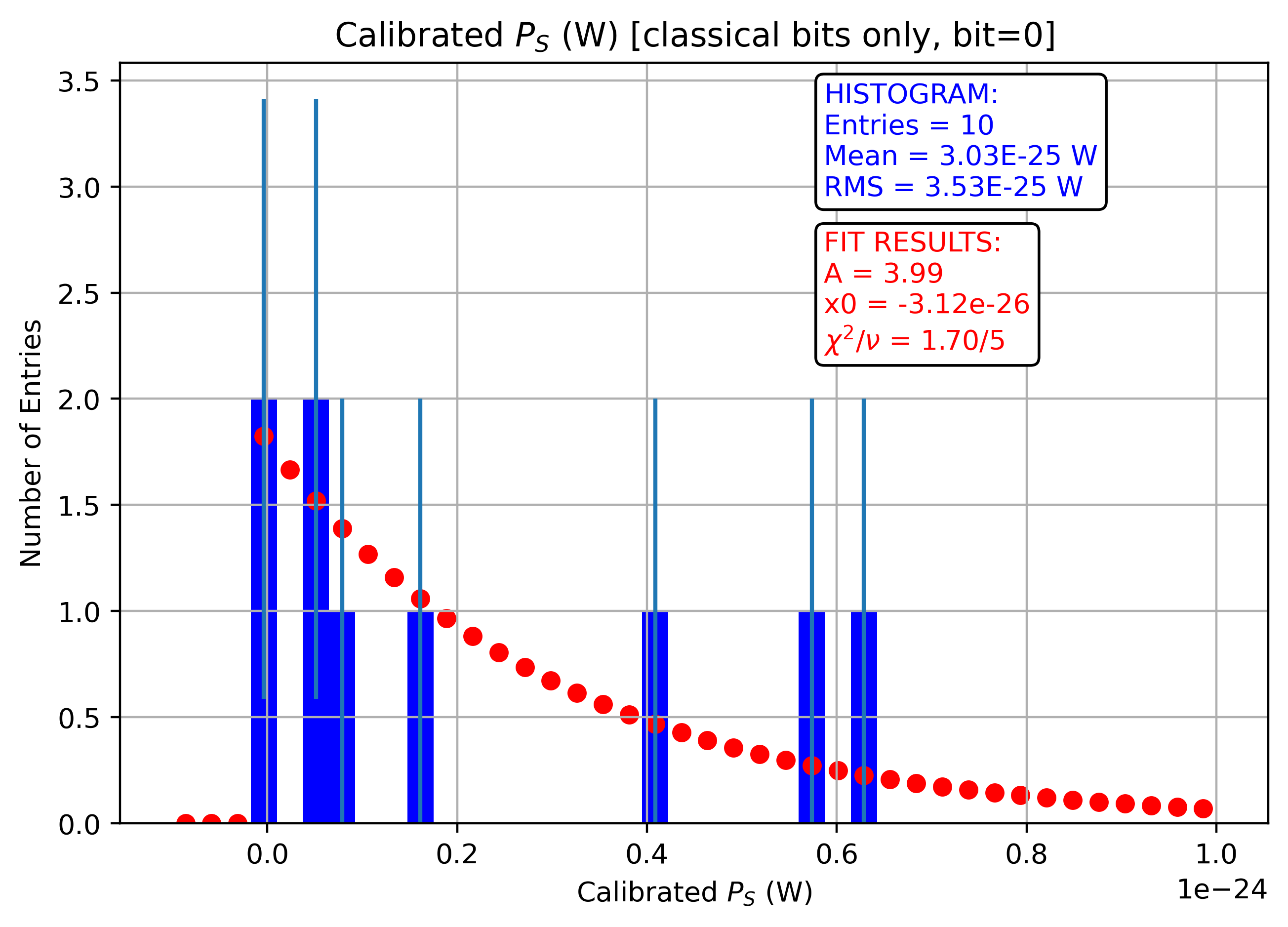}  
    \caption{Calibrated $P_S$ distribution. One entry with negative calibrated power was encountered. }
    \label{fig:calib_power_at_HEMT_input}
\end{figure}

Calibrations described above were applied to raw data collected with the classical sample. The unaveraged noise power spectrum follows a $\chi^2$ distribution with 2 degrees of freedom. The spectrum of the unaveraged calibrated power at the HEMT input, $P_S$, and the fit are shown in Fig.~\ref{fig:calib_power_at_HEMT_input}.  Similar analysis and limit setting has been performed in other experiments that look for power excesses in power spectra without averaging~\cite{PhysRevD.109.083014}. The experiment searches for a power excess at the source frequency in the {bit~=~0} data compared to the background. As described in Ref.~\cite{cervantes2022admx}, in the absence of a power excess, the limit setting procedure sets a 90\% confidence limit using the inverse cumulative function (percent point function) applied to the normal distribution truncated at 0 of the probability of measuring a power excess signal due to NLQM. The power exclusion limit was found at $6.97 \times 10^{-25}$\,W with 90.0\% confidence level (CL). To continue the exercise of deriving a limit on $\epsilon$ using the data obtained with the classical sample, it was necessary to apply several correction factors.

\subsection{Effect of Narrow Frequency Interval}
Choosing 1\,mHz RBW significantly suppresses thermal noise. 
However, some of the NLQM signal leaks outside of this narrow interval. 
Hence, the calibration of $P_A$ needs to be adjusted for this effect. It was found that 85.6\% of the signal was contained inside the 1\,mHz bandwidth. This value was calculated as a ratio of the peak power (maximum power measured in a 1\,mHz-wide interval) divided by the total power measured in the 1\,Hz-wide interval around the peak. In this measurement the signal was propagating from the RF generator to the SA through the complete experiment circuit with the RF switches set in switch 1 port 1, switch 2 port 1 configuration. It was also verified that the ratio does not reduce significantly when the wide interval was extended from 1\,Hz to 2\,Hz. 

\subsection{Fidelity Corrections}
Classical readout error during quantum measurement affects both arms of the wave function. For an event in which such an error occurred (i.e. true quantum bit value 1 was incorrectly observed as 0), the macroscopic data are then recorded for the analysis when they are not supposed to be recorded according to the experiment design. These data do not carry any information about quantum nonlinearity and need to be discarded from the analysis. This cannot be done on the event-by-event basis, hence the effect is taken into account statistically. In case of perfect classical readout fidelity ($F_{C}$), the limit on  $\epsilon_{\gamma}$ would scale as 1/$\sqrt{N}$, where $N$ is the total number of events in the data sample. When the fidelity is less than 100\%, the number of valid events in the data sample is reduced from $N$ to $NF_{C}$, and so the
limit scaling becomes 1/$\sqrt{NF_{C}}$. As a result, the limit needs to be corrected by a multiplicative factor of 1/$\sqrt{F_C}$ = 1.08.

Hadamard gate operation fidelity $F_{H}$  is $| \langle \Phi | \Psi\rangle |^2$, where $|\Psi\rangle$ = (1/$\sqrt{2}$) $(|0\rangle + |1\rangle)$ is our desired state and $ |\Phi\rangle = \alpha |0\rangle + \beta |1\rangle$ is the state to which the qubit is brought instead. By using perturbative expansion of $\alpha$ and $\beta$ in terms of a small correction, setting $\alpha^2 + \beta^2$ = 1 and recalling the definition of $F_{H}$ the correction can be expressed as $\sqrt{1 - F_{H}}$.  Electric field that leaks from one arm to another is $\epsilon_{\gamma} |\alpha|^2 E_0$, hence the multiplicative $F_{H}$ correction on 
$\epsilon_{\gamma}$ can be written as 1/$[2\times(1/2 - \sqrt{(1.0 - F_{H})})]$ = 1.25

The active reset fidelity does not affect the limit derivation: in case of failed active reset, the qubit remains in the excited state, and when the Hadamard gate is applied the qubit returns to an equal superposition of ground and first excited state.

\subsection{Deriving Expected Limit on $\epsilon$ from Classical Data}
All the corrections described above were applied to the power limit. Then the limit on power was converted to the limit on $|\epsilon|$ according to 
Eqn.~(\ref{eqn:epsilon}). The applied power $P_A$ was 7.45\,W.
The fictitious $|\epsilon|$ limit obtained from classical data was $\lessapprox 8.93\times 10^{-13}$ at 90.0\% CL. We reiterate that this is only derived here as a test of the limit setting procedure performed below, and that the classical data cannot test nonlinear quantum dynamics.

\section{\label{sec:qanalysis}Quantum Data Analysis}
The data analysis of the classical bits subset described in the previous section was repeated on the quantum bits data to obtain a calibrated power spectrum. Due to considerations of ``quantum dilution'' safety, this spectrum was never plotted. For the same concerns, the individual power (raw or calibrated) values and the values of the parameters that cumulatively describe the power spectrum, such as the mean and the spread, were maintained blinded during the analysis process. All of this information was only stored in the volatile memory of the computer through the execution of the analysis script. Excess power due to the NLQM signal was searched for by comparing with the same measurement performed with the classical bits. It was required that the mean of the calibrated power in the quantum data be larger than the mean in the classical data plus 5 standard deviations of the classical data. No excess was found. Therefore, a new bound on $|\epsilon|$ was set at $ \lessapprox 1.15 \times 10^{-12}$ at 90.0\% CL.

\section{\label{sec:results}Conclusions}
By using quantum bits generated on the \texttt{Rigetti} \texttt{Aspen-M-3} quantum processor, and spectral measurements of the RF circuit operating at cryogenic temperature, the 90.0\% CL limit was set at $|\epsilon| \lessapprox 1.15 \times 10^{-12}$. This is the most stringent limit on nonlinear quantum mechanics thus far and an improvement by nearly a factor of 50 over the previous experimental limit. This result pushes further the frontier of our current understanding of quantum mechanics.

To further improve experimental sensitivity to nonlinearity effects, in accordance with Eqn.~(\ref{eqn:epsilon}), it is necessary to increase the strength of the generated EM signal, to increase the sensitivity of the detecting instrument, and to improve statistics. In our experiment, the strength of the generated EM signal was increased by using a high-power RF amplifier at room temperature. The detection sensitivity was improved through the suppression of the noise floor of the measurement. This was achieved by moving the essential part of the experimental setup to a cryogenic environment and using a very narrow measurement bandwidth. In this experiment, the data sample size was limited by the total available run time at cryogenic temperature, and the long integration time required for each measurement.

\section{\label{sec:ack}Acknowledgements} 
We thank Florent Lecocq for useful discussions, as well as Fumio Furuta, Alex Netepenko and the Fermilab's APS-TD vertical cavity testing team for support with the facility operation.
This material is based upon work supported by the U.S. Department of Energy, Office of Science, National Quantum Information Science Research Centers, Superconducting Quantum Materials and Systems Center (SQMS) under contract number DE-AC02-07CH11359.

\bibliography{NLQMBibliography}
\end{document}